\documentclass[runningheads]{llncs}
\usepackage{graphicx}
\usepackage{url}
\usepackage{todonotes}
\usepackage{wrapfig,booktabs}
\usepackage{cite}
\usepackage{acronym}
\usepackage{hyperref}
\usepackage{subcaption}
\usepackage{tcolorbox}
\usepackage{orcidlink}

\newcommand\crule[3][black]{\textcolor{#1}{\rule{#2}{#3}}}

\begin{document}

\acrodef{WTR}[WTR]{web table retrieval}
\acrodef{IR}[IR]{information retrieval}
\acrodef{LLM}[LLM]{Large Language Model}
\acrodefplural{LLM}[LLMs]{Large Language Models}
\acrodef{sDCG}[sDCG]{Session-based DCG}
\acrodef{idf}[idf]{inverse document frequency}
\acrodef{CSM}[CSM]{Complex Searcher Model}
\acrodef{SERP}[SERP]{search engine result page}
\acrodef{DCG}[DCG]{discounted cumulated gain}
\acrodef{RBP}[RBP]{rank-biased precision}
\acrodef{sRBP}[sRBP]{Session RBP}
\acrodef{IG}[IG]{information gain}
\acrodef{SERP}[SERP]{search engine result page}
\acrodefplural{SERP}[SERPs]{search engine result pages}

%
\title{Context-Driven Interactive Query Simulations Based on Generative Large Language Models}

\titlerunning{Context-Driven Interactive Query Simulations Based on Generative LLMs}

\author{Björn Engelmann\inst{1}\orcidlink{0009-0000-7074-9066} \and
        Timo Breuer\inst{1}\orcidlink{0000-0002-1765-2449} \and
        Jana Isabelle Friese\inst{2}\orcidlink{0009-0005-2483-0476} \and \\
        Philipp Schaer\inst{1}\orcidlink{0000-0002-8817-4632} \and
        Norbert Fuhr\inst{2}\orcidlink{0000-0002-0441-6949}}

\authorrunning{Engelmann et al.}

\institute{TH Köln - University of Applied Sciences, Germany \\
\email{\{bjoern.engelmann,timo.breuer,philipp.schaer\}@th-koeln.de} \and
University of Duisburg-Essen, Germany \\
\email{\{jana.friese,norbert.fuhr\}@uni-due.de}}

\maketitle              
\begin{abstract}
Simulating user interactions enables a more user-oriented evaluation of \ac{IR} systems. While user simulations are cost-efficient and reproducible, many approaches often lack fidelity regarding real user behavior. Most notably, current user models neglect the user's context, which is the primary driver of perceived relevance and the interactions with the search results. To this end, this work introduces the simulation of context-driven query reformulations. The proposed query generation methods build upon recent \ac{LLM} approaches and consider the user's context throughout the simulation of a search session. Compared to simple context-free query generation approaches, these methods show better effectiveness and allow the simulation of more efficient IR sessions. Similarly, our evaluations consider more interaction context than current session-based measures and reveal interesting complementary insights in addition to the established evaluation protocols. We conclude with directions for future work and provide an entirely open experimental setup.

\end{abstract}
\section{Introduction}

The Cranfield paradigm is the de facto standard approach for evaluating \ac{IR} methods, allowing a fair and reproducible comparison of different retrieval systems. However, these merits come at the cost of a strong abstraction of the user behavior. The underlying user model of Cranfield-style experiments assumes that the user formulates a single query, scans the result list in its entirety until a fixed rank, and judges the relevance irrespective of earlier seen search results. In the real world, users behave differently. 

Simulating user interactions offers a more cost-efficient and reproducible alternative to real-world user experiments. It allows us to conclude  the generalizability of experimental results beyond the boundaries of Cranfield-style experiments. However, current endeavors lack the inclusion of the user's context in the simulations. To this end, our work analyzes the contextual influence on simulated users, focusing on query generation.

To address the changing knowledge state of the user, we propose \textbf{two new query generation methods} based on generative \acp{LLM} that consider different types of context information. The methods leverage this information by incorporating it into the reformulations, thereby enhancing the fidelity of the generated queries to real-world user interactions and improving the quality of the user simulation in total.

In addition, we provide \textbf{a first-of-its-kind comparison of lexical (sparse) and Transformer-based (dense) retrieval methods} for simulated interactive retrieval. While dense retrieval methods typically show higher retrieval effectiveness, they come at a significantly higher cost in terms of time and resources compared to sparse methods. The simulated environment in this work allows for a direct comparison of both approaches in a user-oriented setting.

To assess the results of the simulations, we conduct \textbf{an in-depth evaluation of the simulations}. For a holistic understanding of the considered factors and their effects on retrieval effectiveness, the evaluation includes different perspectives of retrieval effectiveness in simulated interactive retrieval sessions. While prevalent evaluation measures report the information gain during a search session, this work also considers the effort required to acquire that knowledge beyond the reformulation of queries.

Besides the reported findings, we contribute additions for the TREC test collections in the form of new query datasets. Moreover, we provide an \textbf{entirely open and reusable experimental setup}.\footnote{\url{https://github.com/irgroup/SUIR}}

\section{Related Work}

Early experiments with user simulations date back to the 1980s~\cite{DBLP:conf/sigir/TagueNW80,DBLP:conf/sigir/TagueN81}, but recently, the topic got more attention from the \ac{IR} community~\cite{DBLP:journals/sigir/AzzopardiJKS10,DBLP:conf/sigir/BalogM0021}. Like earlier works, we implement the simulations with a user model covering 1) query formulations, 2) scanning of the retrieved lists, 3) selecting and clicking appealing items, 4) reading and judging documents for relevance, and 5) inspecting other items in the result list and making stop decisions either leading to query reformulations or abandoning the search session (cf. Figure~\ref{fig:user_model}). The literature provides different simulation frameworks~\cite{DBLP:conf/ictir/CarteretteBZ15,DBLP:conf/cikm/MaxwellAJK15,DBLP:journals/ir/PaakkonenKKAMJ17,DBLP:conf/ictir/ZhangLZ17}. However, this work mainly aligns with the \textit{Complex Searcher Model} implemented by the \texttt{SimIIR} toolkit~\cite{DBLP:conf/sigir/MaxwellA16,DBLP:conf/cikm/Zerhoudi0PBSHG22}. 

Generating high-fidelity queries is an important aspect of interactive user simulations~\cite{DBLP:conf/ictir/CarteretteBZ15,DBLP:conf/cikm/BaskayaKJ13,gunther2021assessing,DBLP:conf/ecir/BreuerFS22}. Earlier works propose different query generation approaches that either rely on principled rules or language models. In recent times, there has been a notable focus on  leveraging LLMs for query generation~\cite{Mackie2023,Wang2023Query2docQE,wang2023generative}. Doc2Query is a method that produces a set of questions that a given document may answer.
The first presented use of these questions was to enrich documents during indexing \cite{DBLP:journals/corr/abs-1904-08375}. This enrichment significantly increased the retrieval effectiveness for term-based ranking methods like BM25. \cite{10.1145/3583780.3615187} have shown that this method is also suitable for varying queries in a simulation environment.
Recently, instruction-tuned \acp{LLM} were used to generate queries based on a given prompt providing context information about the information need~\cite{DBLP:conf/sigir/AlaofiGSS023}. This approach has not been used for simulating interactive \ac{IR} sessions, which is a contribution of this work.

A commonly used measure to simulate sessions with multiple query reformulations is the \ac{sDCG}~\cite{DBLP:conf/ecir/JarvelinPDN08}, which adds an additional discount to the \ac{DCG}~\cite{DBLP:conf/sigir/JarvelinK00} by the query position in the session. More recently, a session-based variant of  \ac{RBP}~\cite{DBLP:journals/tois/MoffatZ08} that also builds upon a query-based discount was introduced as \ac{sRBP}~\cite{DBLP:conf/ictir/LipaniCY19}. The experiments by Lipani et al.~\cite{DBLP:conf/ictir/LipaniCY19} suggest that sDCG and sRBP lead to different outcomes and let the authors conclude that both measures provide different evaluation perspectives. For a more in-depth review of user simulations, we refer the reader to Balog and Zhai~\cite{DBLP:journals/corr/abs-2306-08550}. 

\section{Methodology}

\begin{figure}[!t]
    \centering
    \includegraphics[width=\textwidth]{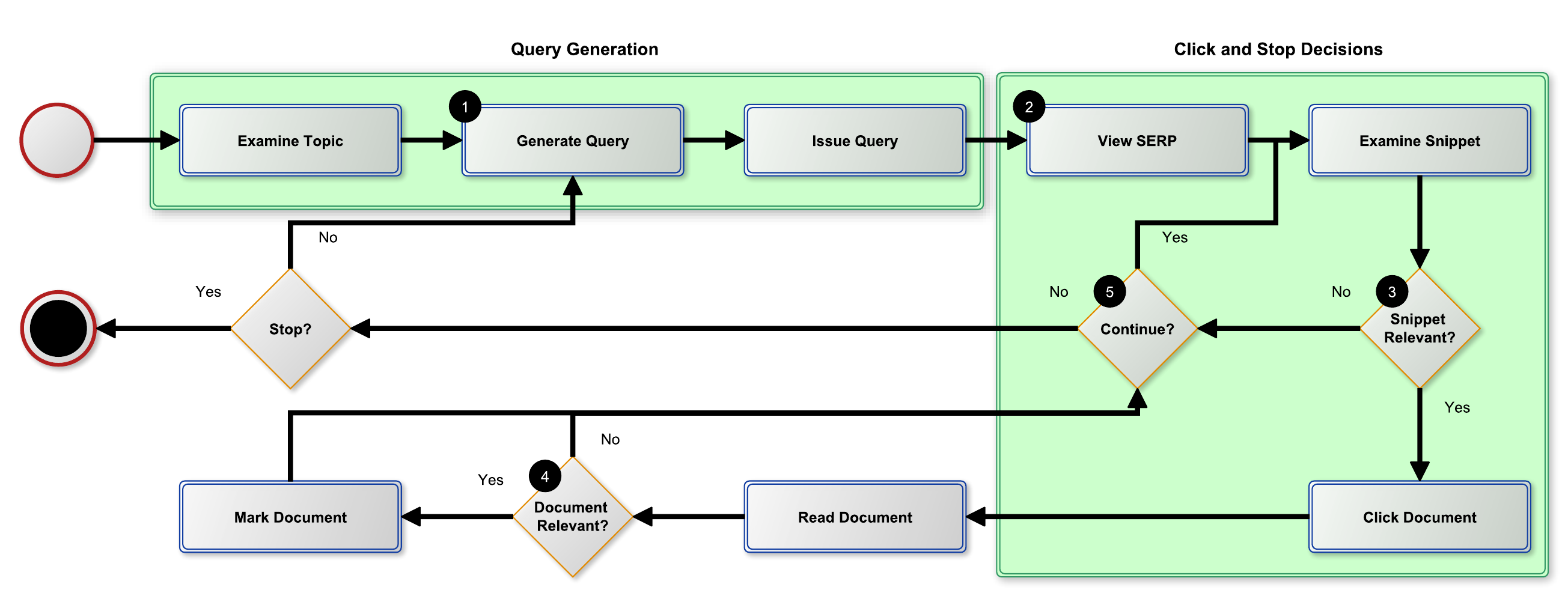}
    \caption{Focus of this work in the context of the Complex Searcher Model~\cite{DBLP:conf/cikm/MaxwellAJK15}.}
    \label{fig:user_model}
\end{figure}

Similar to earlier works~\cite{DBLP:conf/ictir/CarteretteBZ15,DBLP:conf/cikm/MaxwellAJK15,DBLP:journals/ir/PaakkonenKKAMJ17,DBLP:conf/ictir/ZhangLZ17}, we ground our simulations on a defined sequence of interactions with the retrieval system and its search results, visualized in Figure~\ref{fig:user_model}. In this work, the particular focus is the \textbf{query generation} and the \textbf{interaction with \acp{SERP}} in different \textit{contexts}. 
We evaluate simulated sessions under two different retrieval paradigms: sparse, i.e., lexical-based ranking methods, and dense, i.e., Transformer-based methods (cf.~\ref{sec:retrieval_models}).
Regarding the interaction with \acp{SERP}, we analyze user-specific click decisions and context-dependent stopping strategies (cf.~\ref{sec:click_stop_decisions}). 
Additionally, we analyze the user's \textit{scope of the information need} and the implications of \textit{interactive feedback} for query reformulations  (cf.~\ref{sec:query_generation}). The simulation runs we examine are analyzed with three different measures (cf.~\ref{sec:eval_measures}).

\subsection{Retrieval Models}
\label{sec:retrieval_models}

\begin{table}[!t]
    \centering
    \caption{Adhoc retrieval effectiveness of BM25 and BM25+MonoT5 evaluated with the New York Times Annotated Corpus (\textbf{Core17}) and the TREC Washington Post Corpus (\textbf{Core18}) that are part of the experimental setup (cf. \ref{sec:implementation}). The cutoff values denote the number of documents reranked by MonoT5.}
    \resizebox{\textwidth}{!}{
    \begin{tabular}{l|c|c|c||c|c|c}
            \toprule
            \multicolumn{1}{c}{} & \multicolumn{3}{c}{\textbf{Core17}} & \multicolumn{3}{c}{\textbf{Core18}} \\
            \midrule
            Model & \hspace*{1.5mm} P@10 \hspace*{1.5mm} &  nDCG@10 & \hspace*{1.5mm} Bpref \hspace*{1.5mm} & \hspace*{1.5mm} P@10 \hspace*{1.5mm} &  nDCG@10 & \hspace*{1.5mm} Bpref \hspace*{1.5mm}\\    
            \midrule
            BM25            &	0.458&	0.372&	0.216 & 0.404 & 0.371 & 0.222 \\
        	MonoT5@50       &	0.562&	0.451&	0.139 & 0.480 & 0.454 & 0.195\\
        	MonoT5@100      &	0.558&	0.454&	0.185 & 0.482 & 0.449 & 0.232\\
        	MonoT5@200      &	0.574&	0.468&	0.232 & 0.476 & 0.447 & 0.261\\
            MonoT5@500      &	0.566&	0.460&	0.283 & 0.460 & 0.435 & 0.292\\
            MonoT5@1000     &	0.550&	0.447&	0.312 & 0.466 & 0.439 & 0.311\\
            \bottomrule
    \end{tabular}
    }
\label{tab:bm25_vs_monot5}
\end{table}

The retrieval model is a key element in our simulations. In the experiments, we evaluate the benefits of using a more effective model than a baseline ranking. To this end, we compare BM25~\cite{DBLP:journals/ftir/RobertsonZ09} against MonoT5~\cite{DBLP:conf/emnlp/NogueiraJPL20} that is known to outperform the former method in a zero-shot setting. To better understand the quality of search results shown to the simulated users, we evaluate the retrieval effectiveness in a preliminary test collection-based adhoc retrieval experiment. 

Table~\ref{tab:bm25_vs_monot5} compares BM25 to MonoT5 at different cutoffs, i.e., the cutoff level determines how many documents are retrieved by the first-stage ranker and will be re-ranked by MonoT5. 
It is important to emphasize that MonoT5 works according to the retrieve and rerank paradigm, which means that in our case, a set of documents is retrieved in the first phase with a sparse model (BM25 in our case) and reranked in a second phase by a dense model.
In principle, the re-ranker can rely on a larger set of potentially relevant documents and place them at higher ranks. 
However, an increasing cutoff level does not substantially improve P@10 and nDCG@10, which can be explained by unjudged documents placed at higher ranks, as also shown by the increasing Bpref scores of MonoT5.

In our simulations, we want to compare BM25 against a strong competitor that is also computationally efficient, as the user simulations imply several other computations as well. For this reason, we consider MonoT5@100 to be sufficient for our experiments. A lower cutoff level substantially reduces the computation time and also helps to reduce the environmental impact of our research~\cite{DBLP:conf/sigir/ScellsZZ22}. Higher cutoff levels do not lead to remarkable improvements in precision, which is of primary interest for our simulation experiments.

From a more general perspective, these evaluations further reveal the limitations of using ad-hoc test collections for simulated interactive retrieval experiments. When there are many unjudged documents in the rankings, the simulated users are indecisive about clicks, and in this case, the true potential of Transformer-based models cannot be fully analyzed in simulations when the test collection is biased towards a particular kind of retrieval method. We consider it a part of the future work to provide adequate resources, data, and tools in this regard.

\subsection{Click and Stop Decisions}
\label{sec:click_stop_decisions}

\begin{table}[!t]
    \caption{Overview of our user simulation configurations based on query generation methods and click models.}
    \begin{minipage}{.5\linewidth}
      \centering
      \subcaption{Query generation methods}
      \label{tab:query_methods}
        \resizebox{.8\textwidth}{!}{
        \begin{tabular}{l|c|c|c}
        \toprule
        Strategy & Topic & Feedback & Generation \\
        \hline
        GPT & - & - & Probabilistic \\
        GPT+ & + & - & Probabilistic \\
        GPT* & + & - & Rule-based\\
        GPT** & - & - & Rule-based\\
        D2Q & - & - & Rule-based\\
        D2Q+ & - & + & Rule-based\\
        D2Q++ & + & + & Rule-based \\
        \bottomrule
        \end{tabular}
        }
      \medskip
    \end{minipage}
    \begin{minipage}{.5\linewidth}

      \centering
    
      \subcaption{Click models}
      \label{tab:click_model}
      \resizebox{0.65\textwidth}{!}{
        \begin{tabular}{l|c|c}
        \toprule
        Model & \hspace*{1mm} \textit{rel} \hspace*{1mm} & \hspace*{1mm} \textit{nrel} \hspace*{1mm} \\
        \hline
        Perfect & 1.0 & 0.0 \\
        Navigational & 0.9 & 0.1 \\
        Informational & 0.8 & 0.4 \\
        Almost random & 0.6 & 0.4 \\
        \bottomrule
        \end{tabular}
        }
      \medskip
    \end{minipage}
\end{table}

To be in line with earlier work~\cite{DBLP:conf/wsdm/HofmannSWR13}, we compare four different types of click behavior in our experiments. All of the methods are based on probabilistic decisions biased towards the level of relevance and are implemented as shown in Table~\ref{tab:click_model}. We note that other ways exist to simulate clicks~\cite{10.1145/3623640}, when there are historical interaction logs, which unfortunately were not available for our experimental setup. The \textit{perfect} and \textit{almost random} browsing behaviors serve as sky- and baselines, respectively. The \textit{perfect} user always clicks on relevant results and never clicks non-relevant or unjudged documents, i.e., the user \textit{scents} the relevant information in a strongly idealistic way. In contrast, the \textit{almost random} behavior mainly neglects the relevance labels. In between, there are the \textit{informational} and \textit{navigational} models with different degrees of exploratory behavior. 
We model the stop decisions in two different ways. 

\textbf{Static} The first stopping criterion is based on the naive assumption that the user browses a fixed number of \textit{results per page (rpp)} and then continues the session by reformulating another query about the topic. 

\textbf{Dynamic} In addition, we implement a time-based stopping criterion, first applied by Maxwell~\cite{DBLP:phd/ethos/Maxwell19} in simulated \ac{IR} sessions, that is based on the \textit{give-up} heuristic by Krebs et al.~\cite{Krebs1974} stemming from the foraging theory. More precisely, the simulated user has a fixed time budget (\textit{tnr}) that starts to deplete after the last seen relevant search result. If another relevant document is seen, the time budget resets. If the budget is depleted, the user reformulates the query.

\subsection{Query Generation}
\label{sec:query_generation}

In the following, we describe our query generation methods, for which an overview is given in Table~\ref{tab:query_methods}. Generally, we employ two different \textit{strategies}: prompting GPT to output query strings and summarizing document contents with the help of Doc2Query (D2Q). Depending on the simulated context, we provide the simulated user with the contents of the \textit{topic} file, i.e., the description and narrative, and let the user consider the \textit{feedback} of earlier seen search results. On the one side, the generation methods can be classified as fully \textit{probabilistic} since they rely entirely on the language models' outputs. Conversely, there are \textit{rule-based} methods that expand the query stem based on the topic's title with newly generated query terms in a principled way.
    
\subsubsection{Prompting query reformulations}

Given the information need described by the topic file, we construct a topic-specific prompt for the LLM. More specifically, we let the LLM generate outputs based on the prompt template given below. 

\begin{tcolorbox}[title=\textbf{Prompt template for the query strategies \textbf{``GPT''} and  \textbf{``GPT+''}},size=small,fontupper=\small, fontlower=\small]
  \texttt{\textcolor{magenta}{Please generate one-hundred keyword queries about \textbf{<title>}}. \textcolor{cyan}{\textbf{<description>} \textbf{<narrative>}}}
\end{tcolorbox}

\texttt{<title>}, \texttt{<description>}, and \texttt{<narrative>} are taken from the respective fields of the particular topic. To better understand the effects of different prompting strategies, we consider the first generation method \textbf{GPT} (\crule[magenta]{.3cm}{.3cm}) that omits the additional topic fields \texttt{<description>} and \texttt{<narrative>} in the prompt to have context-free query variants. The second method \textbf{GPT+} (\crule[magenta]{.3cm}{.3cm}\crule[cyan]{.3cm}{.3cm}) makes use of the entire prompt to generate query reformulations. 

We implement two additional methods to better understand the benefits of having different query variants with semantics. \textbf{GPT*} uses the topic's title as the seed query for each query (re)formulation and expands it with a single term out of the vocabulary generated for the topic by the \textbf{GPT+} (\crule[magenta]{.3cm}{.3cm}\crule[cyan]{.3cm}{.3cm}) strategy. Similarly, \textbf{GPT**} is also a rule-based variant based on the context-free vocabulary of \textbf{GPT} (\crule[magenta]{.3cm}{.3cm}). For each topic in the test collections (cf.~\ref{sec:implementation}), we adapt the prompting strategy to query OpenAI’s API and parse the outputs of GPT-3.5 (more specifically, gpt-3.5-turbo-0301\footnote{\url{https://platform.openai.com/docs/model-index-for-researchers}}). In total, we generated 400 queries for all 50 topics in two test collections, which resulted in a total of 40,000 generated queries that stem from the outputs of OpenAI's LLM.

\subsubsection{Doc2Query}
Doc2Query relies on a set of documents to generate queries. To obtain the first set of documents, a seed query is required, which equals the \texttt{<title>} of the corresponding topic in this case. 
Our Doc2Query generation approach is based on \cite{10.1145/3583780.3615187} and extends the methodology by integrating background information.
Unlike the GPT variants, this method allows to generate queries dynamically at runtime, taking seen results and, therefore, the context of the current session into account. To factor in the user’s changing knowledge state, terms are extracted from seen documents and added to the current knowledge state. This approach assumes that terms that come from documents actually seen are more likely to satisfy the information need than random terms from the corpus. For each query reformulation, a term is taken from the knowledge state and added to the seed query.
The knowledge state at the time of the $i$-th query $KS_i$ is a set of terms defined according to: 
\begin{equation}
    \label{eq:knowledge_state}
    KS_i = \bigcup_{j \in \{1,\dots,i\}} \phi(\theta(Q_j)).
\end{equation}
$\theta(Q_j)$ is the set of seen documents for the $j$-th query.
The terms extracted for a given set of documents are defined by: 
\begin{equation}
    \label{eq:keywords}
    \phi(D) = \bigcup_{d \in D} \{t \in d2q(d) \,|\, idf(t) < 0.5 \land t \notin S \}.
\end{equation}
All terms identified by the Doc2Query function $d2q$ are joined, except for those with an inverse document frequency (IDF) below $0.5$ and stopwords.

To determine the effect of background information and relevance feedback on cumulative retrieval effectiveness, we present three different Doc2Query variants.
\textbf{D2Q} serves as the baseline of the Doc2Query variants, which generates queries without background information and feedback.
Here, the knowledge state is based solely on terms taken from seen documents. 
\textbf{D2Q+} integrates the relevance feedback of the simulated user. In addition to $KS_i$, a knowledge state $KS_{i}^{rel}$ is used, which only contains terms from seen documents that have been marked as relevant by the simulated user. As long as $KS_{i}^{rel}$ is not empty, \textit{D2Q+} uses these terms to generate queries. Otherwise, terms from $KS_{i}$ are used:
\begin{equation}
\label{eq:query_exp}
  Q_{i+1} = \left\{\begin{array}{llllll}
        Q_0 \cup t^{rel} & : & t^{rel} & \in & KS_{i}^{rel} & KS_{i}^{rel} \neq \emptyset  \\ 
        Q_0 \cup t & : & t & \in & KS_{i}  & else.
        
        \end{array} \right .
\end{equation}
\textbf{D2Q++} extends the \textit{D2Q+} mechanism with background information. For this purpose, a set of terms is generated from the \texttt{description} and \texttt{narrative} fields. Because these terms are topic-specific, they add another form of context to the query generation. The topic-specific terms are used until no more new queries can be generated, and then the \textit{D2Q+} method takes over.

\subsection{Evaluation Measures}
\label{sec:eval_measures}

\subsubsection{Effort vs. Effect}

At the most granular level, we evaluate the effect of a search session by determining the cumulated \ac{IG} over the costs of all logged interactions, i.e., the effort, by:

\begin{equation}
    \mathrm{Effect} = \sum_{s \in \mathcal{S}} \mathrm{IG} (s), \qquad \mathrm{IG} (s) = \left\{\begin{array}{ll}
        \mathrm{rel}_d, & \mathrm{if} \ s = s_{\mathrm{rel}} \\ 
        0, & \mathrm{else}.
        \end{array} \right .
\end{equation}

$\mathrm{rel}_d$ is the relevance level of the document $d$, $\mathcal{S}$ is the set of all logged interactions, and $s_{rel}$ denotes the particular session log considering the read document as relevant. We note that in this case, the costs are modeled by several additional factors besides the query formulations, including the click decision based on the snippet, the reading of the corresponding document, and making a judgment of the document's relevance for a given topic, and there is no discount over the progress of the session. For the costs of each action, we use the default values of \texttt{SimIIR}.

\subsubsection{Session-based DCG}

As proposed in earlier work, we evaluate the simulated sessions by the \ac{sDCG}~\cite{DBLP:conf/ecir/JarvelinPDN08}, which shows high correlations with real user effectiveness~\cite{DBLP:conf/airs/HagenMS16}. \ac{sDCG} discounts the cumulated gain document- and also query-wise by the logarithm and the corresponding base $bq$ as well as by the query position $i$ in a session:

\begin{equation}
    \mathrm{sDCG} = \sum_{i \in \{1,\dots,n_j\}} \frac{\mathrm{DCG}_{q_i}}{1 + \log_{bq}(i)}, \qquad \mathrm{DCG}_{q_i}=\sum_{r \in \{1,\dots,n_d\}}\frac{2^{\mathrm{rel}_r}-1}{\log_2(r+1)},
\end{equation}

where $n_j$ denotes the number of available queries for a topic $j$ and $bq$ is a free parameter set to $4$. The \ac{DCG}~\cite{DBLP:conf/sigir/JarvelinK00} is defined by the log-harmonic discounted sum of relevance $\mathrm{rel}_r$ of $n_d$ documents in the ranking corresponding to query $q_i$. Obviously, \ac{sDCG} exclusively considers query formulations as \textit{costs} and models the user's stopping behavior by a log-harmonic probability distribution over the number of queries and documents.

\subsubsection{Session RBP}

Similar to \ac{sDCG}, there is a session-based variant of \ac{RBP}~\cite{DBLP:journals/tois/MoffatZ08}, which was introduced by Lipani et al.~\cite{DBLP:conf/ictir/LipaniCY19} and is defined as follows:

\begin{equation}
    \mathrm{sRBP}=(1-p)\sum_{i \in \{1,\dots,n_j\}}\left(\frac{p-bp}{1-bp}\right)^{i-1}\sum_{r \in \{1,\dots,n_d\}}(bp)^{r-1}\cdot \mathrm{rel}_{r},
\end{equation}

where $p$ models the user's persistence similar to \ac{RBP} and $b$ is introduced as a parameter that balances reformulating a query and continuing to browse the result list. \ac{sRBP} discounts later results more than \ac{sDCG}. Because of the considerable length of the simulated search sessions, we set $p$ to $0.99$ to ensure that results found later in the sessions would still contribute to the outcome. For a good trade-off between results later in the queries and results from later queries, we set $b$ to $0.9$. Like the comparison between \ac{DCG} and \ac{RBP}, evaluating \ac{sDCG} and \ac{sRBP} provides two different perspectives of retrieval effectiveness that can be explained by the underlying user models based on different probability distributions.

\subsection{Implementation Details and Datasets}
\label{sec:implementation}

We implement the experimental setup with a rich set of state-of-the-art software libraries, including \texttt{Pyterrier}~\cite{DBLP:conf/cikm/MacdonaldTMO21}, \texttt{ir\_datasets}~\cite{DBLP:conf/sigir/MacAvaneyYFDCG21}, \texttt{SIMIIR v2.0}~\cite{DBLP:conf/cikm/Zerhoudi0PBSHG22}, and HuggingFace's \texttt{transformers}~\cite{DBLP:conf/emnlp/WolfDSCDMCRLFDS20}. We ground the experiments on two TREC newswire test collections, i.e., the New York Times Annotated Corpus\footnote{\url{https://catalog.ldc.upenn.edu/LDC2008T19}} used as part of TREC Common Core 2017 (Core17)~\cite{DBLP:conf/trec/AllanHKLGV17}, and the TREC Washington Post Corpus\footnote{\url{https://trec.nist.gov/data/wapost/}} used as part of TREC Common Core 2018 (Core18)~\cite{DBLP:conf/trec/2018}. Here it is important to note that Core18 includes $\sim$26k ($\sim$4k positive) relevance labels, while Core17 includes $\sim$30k ($\sim$9k positive) labels.
For generating queries with Doc2Query, we use the available pre-trained model without finetuning \footnote{\url{https://github.com/terrierteam/pyterrier_doc2query}}.
All of the experiments are run on a Dell workstation with an Intel i9-12900K CPU, 64GB of RAM, and an NVIDIA GeForce RTX 3070 GPU on Ubuntu 22.04 LTS.

\section{Experimental Results}
The following experiments investigate how the information gain develops for increasing costs. In our case, the cost is either the number of queries or the time spent, expressed in time units.
The evaluations of the simulation runs are based on different user behaviors regarding click and stop decisions (cf.~\ref{sec:click_stop_decisions}), two retrieval systems (cf.~\ref{sec:retrieval_models}), and seven different methods for varying queries (cf.~\ref{sec:query_generation}).
Three measures (cf.~\ref{sec:eval_measures}) are evaluated, providing different perspectives on cost-benefit trade-offs and using two different test collections. 
To determine the effect of specific methods on the information gain, a default configuration is set for each method type and varied only over the aspect that is being evaluated. This default configuration is: \textit{BM25} - \textit{GPT} - \textit{informational} - \textit{10rpp}. 
In our study, we evaluate the differences between the various configurations in an exploratory manner. To make statements about the effectiveness of the different configurations, we compare both their order and the trend of the information gain pairwise over the simulated runs.

\subsection{Retrieval Models and Users}
\label{subsec:dense_sparse_eval}

\begin{figure*}[!t]
\centering
\includegraphics[width=.325\textwidth]{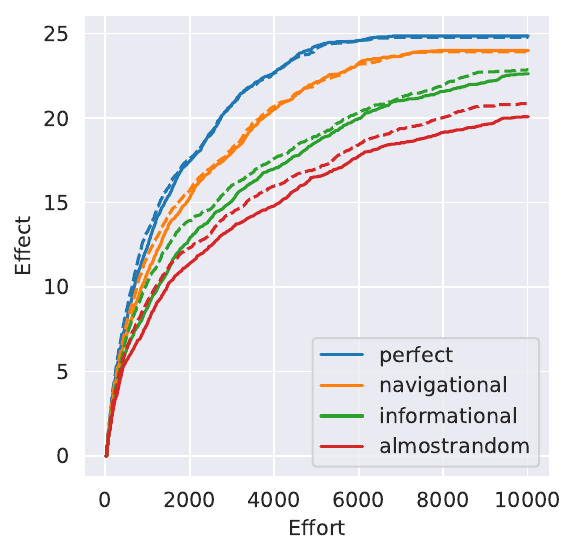}
\includegraphics[width=.325\textwidth]{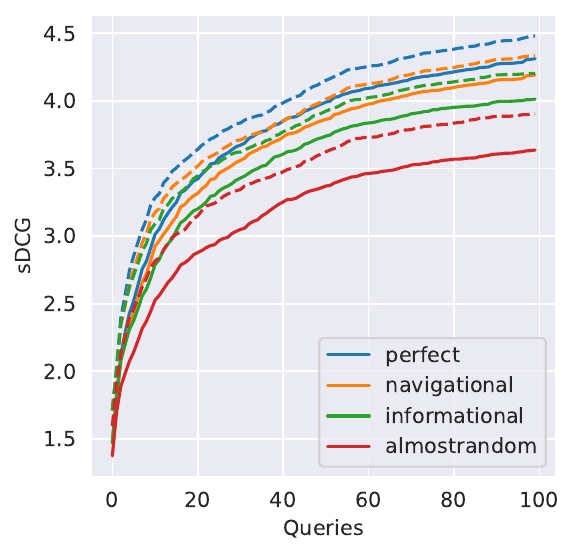}
\includegraphics[width=.325\textwidth]{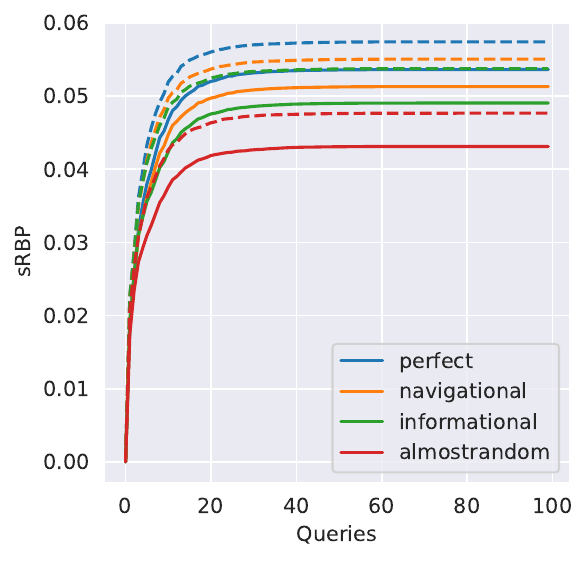}
\includegraphics[width=.325\textwidth]{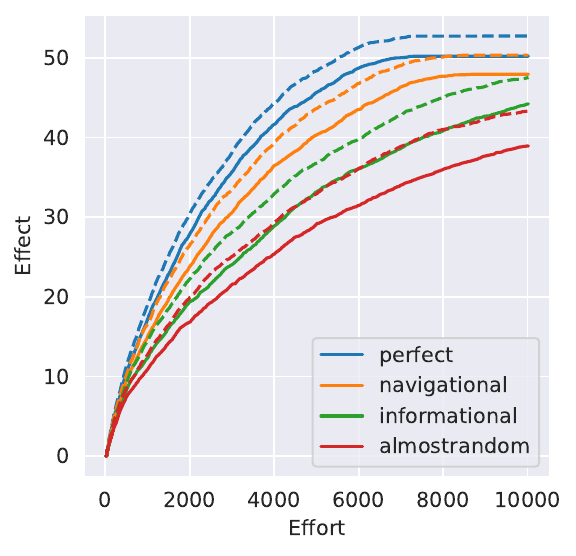}
\includegraphics[width=.325\textwidth]{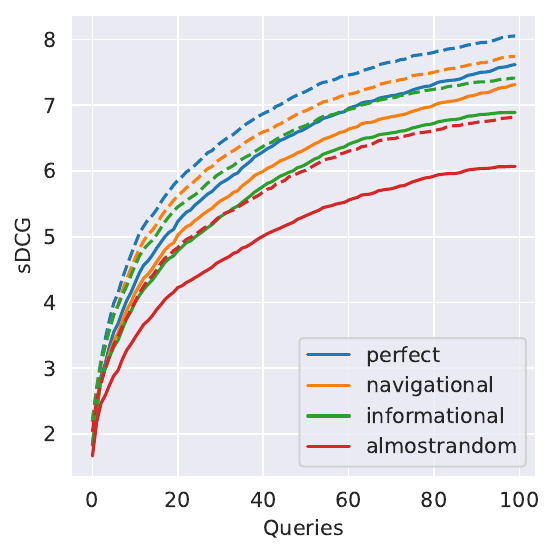}
\includegraphics[width=.325\textwidth]{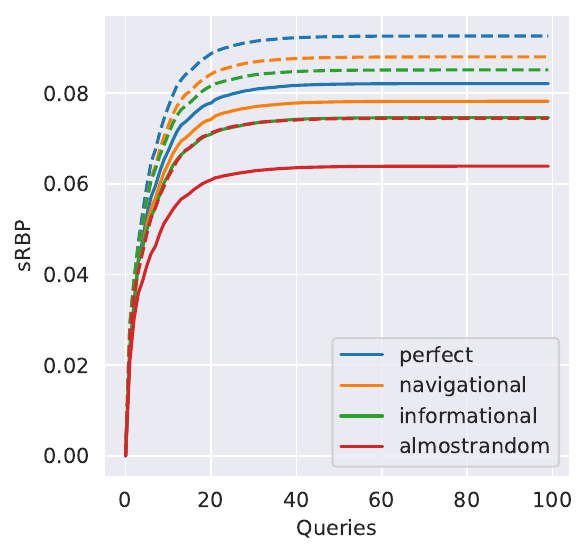}
\caption{Results of the simulated sessions for different click behaviors. Core18 (top) and Core17 (bottom). BM25 in solid lines, MonoT5 dashed, respectively.}
\label{fig:experimental_results_click_decs}
\end{figure*}

This section investigates the influence of the retrieval system and different types of simulated users.
Both user characteristics, i.e., the click behavior (cf.~\ref{tab:click_model}) and the stop decisions, are evaluated with BM25 and MonoT5.
In Figure~\ref{fig:experimental_results_click_decs}, we see a clear order if we look at the sDCG values for the different click profiles. The closer the user is to the perfect click behavior, the higher their information gain. 
Comparing all pairs of retrieval models over the click profiles (e.g., BM25-perfect  vs. MonoT5-perfect) shows a slight dominance of MonoT5.
This effect corresponds to the results of the ad hoc retrieval from Table~\ref{tab:bm25_vs_monot5}. Compared to the sDCG plots, the effort-based evaluation shows a clear difference in user behavior, while the retrieval model plays a lesser role.

\begin{figure*}[!t]
\centering
\includegraphics[width=.325\textwidth]{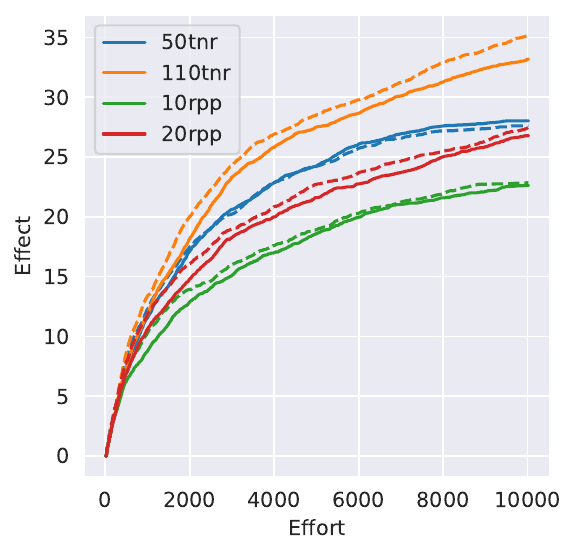}
\includegraphics[width=.325\textwidth]{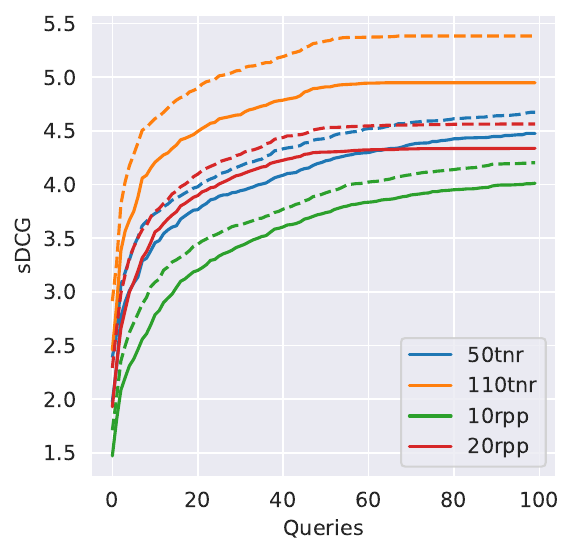}
\includegraphics[width=.325\textwidth]{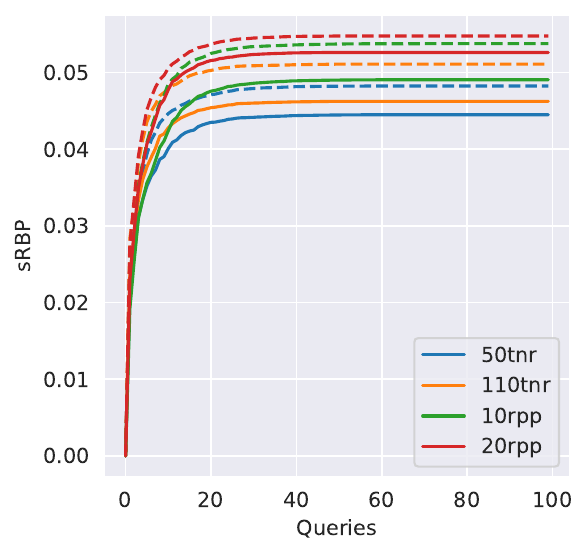}
\includegraphics[width=.325\textwidth]{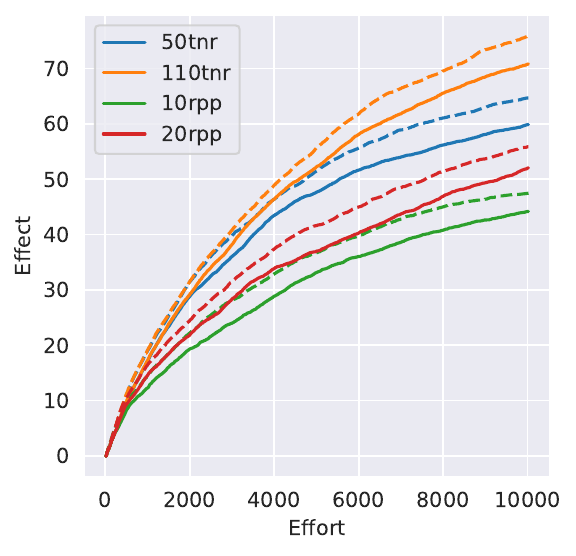}
\includegraphics[width=.325\textwidth]{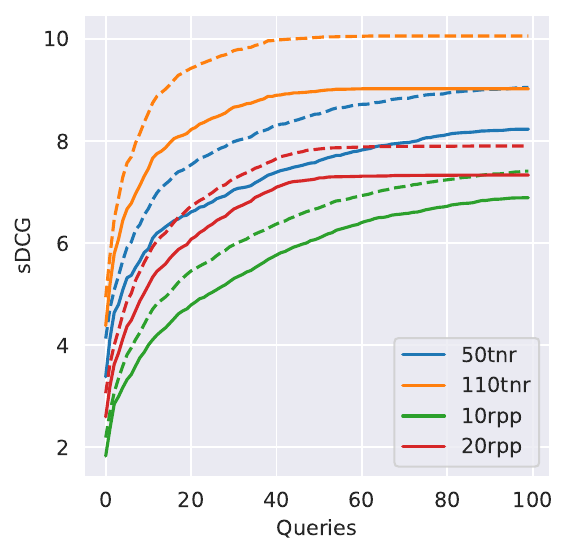}
\includegraphics[width=.325\textwidth]{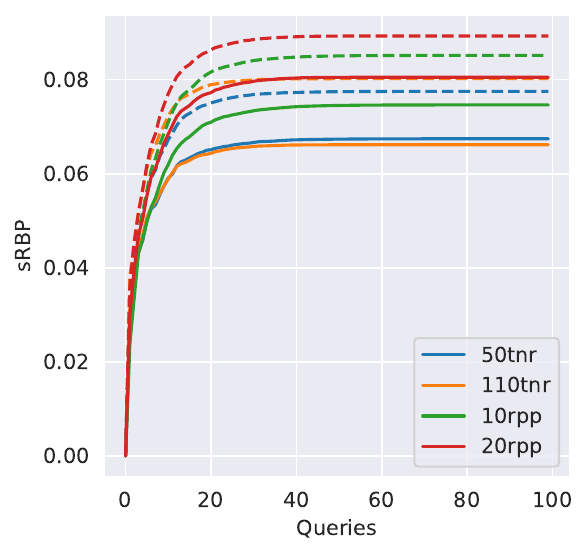}

\caption{Results of the simulated sessions for different stop decisions. Core18 (top) and Core17 (bottom). BM25 in solid lines, MonoT5 dashed, respectively.}
\label{fig:experimental_results_stop_decs}
\end{figure*}

We distinguish two types of stop decisions (cf.~\ref{sec:click_stop_decisions}).
Either a new query is formulated when a fixed number of snippets has been considered (\textit{rpp}) or when a defined time has passed since the last relevant document was examined (\textit{tnr}). 
Each type is evaluated with two parameters: $tnr \in \{50, 110\}$ and $rpp \in \{10, 20\}$.
In Figure~\ref{fig:experimental_results_stop_decs}, it can be seen that the dynamic click behavior positively affects the information gain.
In addition, both an increase in the number of documents examined and an increase in the dynamic time budget have a positive effect.
Again, the effort-based measure weights the impact of user behavior more heavily than the effectiveness of the retrieval model. Furthermore, sRBP is shown to punish dynamic stop behavior.

\begin{figure*}[!t]
\centering
\resizebox{\textwidth}{!}{
\includegraphics[width=.49\textwidth]{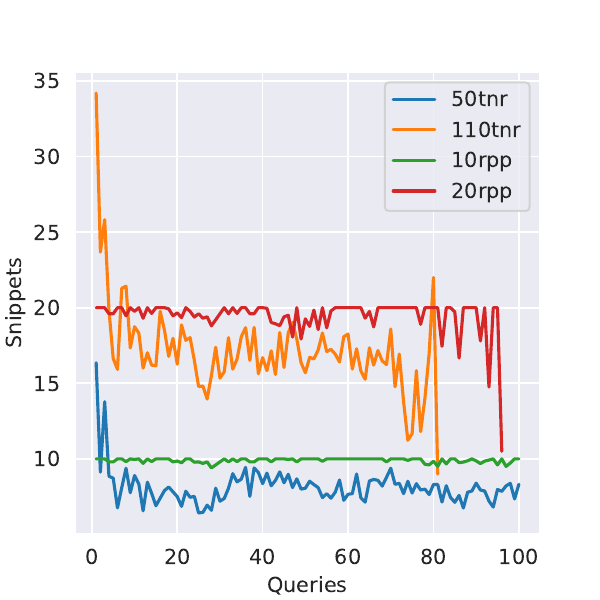}
\includegraphics[width=.49\textwidth]{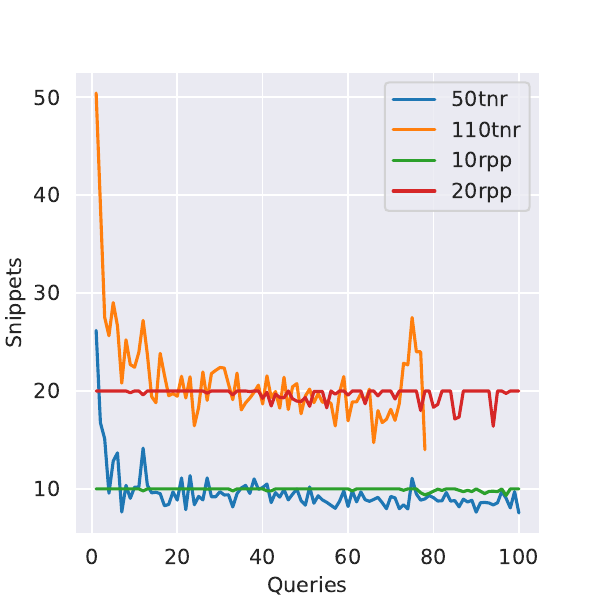}
}
\caption{Distributions for the average number of snippets examined for $i$-th query. Core18 (left) and Core17 (right).}
\label{fig:snippet_dist}
\end{figure*}

To put the different stop profiles in an insightful context, Figure~\ref{fig:snippet_dist} shows the distribution over the average number of snippets considered across the issued queries. First, it is noticeable that both runs of \textit{rpp} vary much less than the \textit{tnr} variants. This effect was expected since the dynamic criterion's stop decision depends on the relevant documents in the result list. In contrast, the upper limit is strictly fixed for the static criterion, and downward outliers only occur when the result list is smaller than the fixed value or the global time limit is reached.
Furthermore, the dynamic variants examine more snippets on average at the beginning of a session and fewer snippets during the session than the corresponding static variants. For \textit{110nr}, a termination can be seen at about 80 queries. This is because the global time budget is reached here. 

\subsection{Query Variation}
\begin{figure*}[!t]
\centering

\includegraphics[width=.325\textwidth]{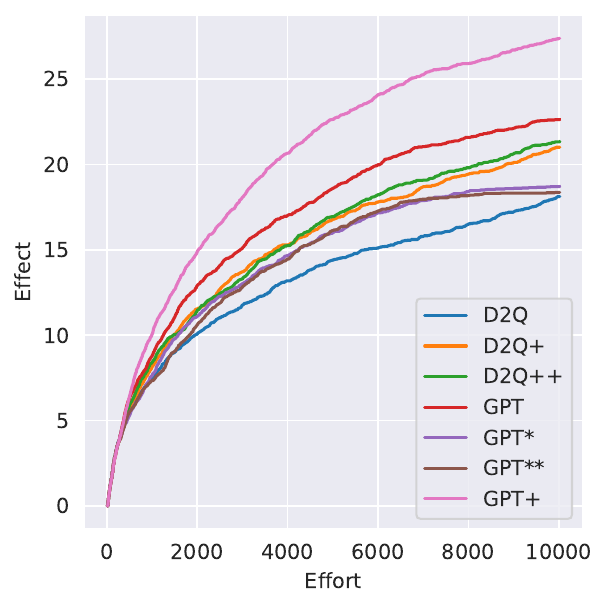}
\includegraphics[width=.325\textwidth]{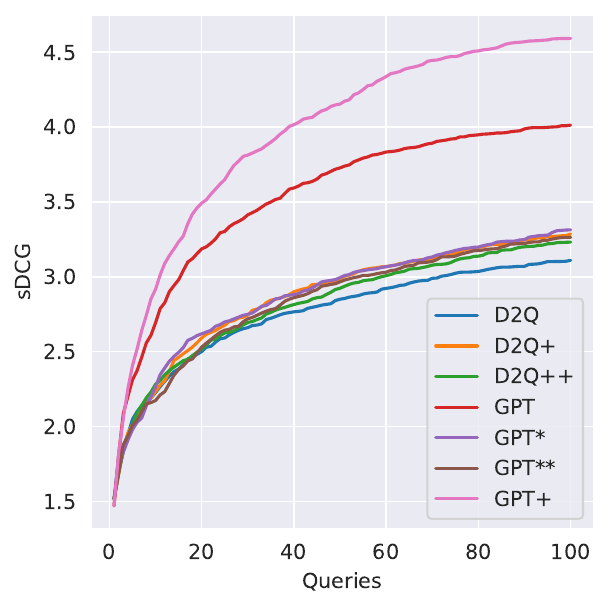}
\includegraphics[width=.325\textwidth]{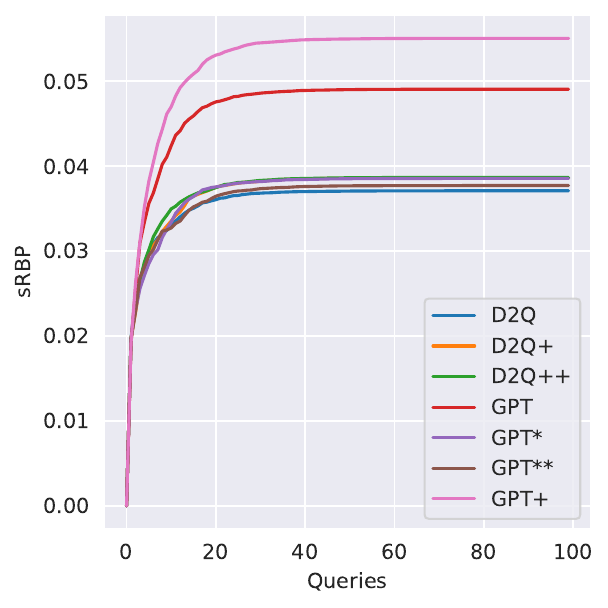}
\includegraphics[width=.325\textwidth]{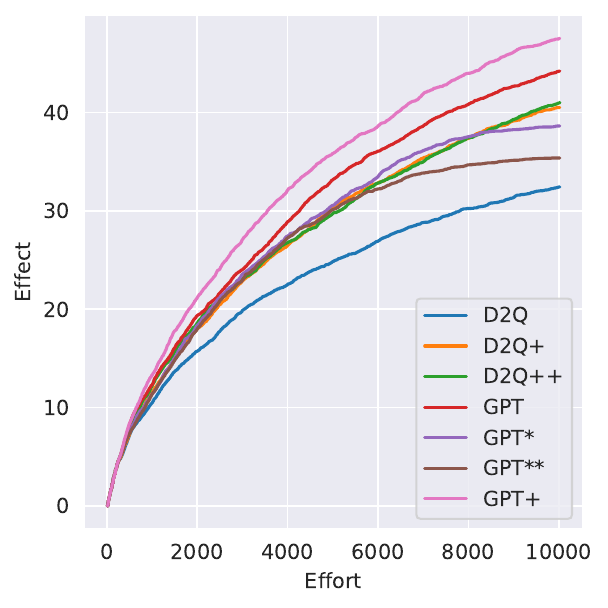}
\includegraphics[width=.325\textwidth]{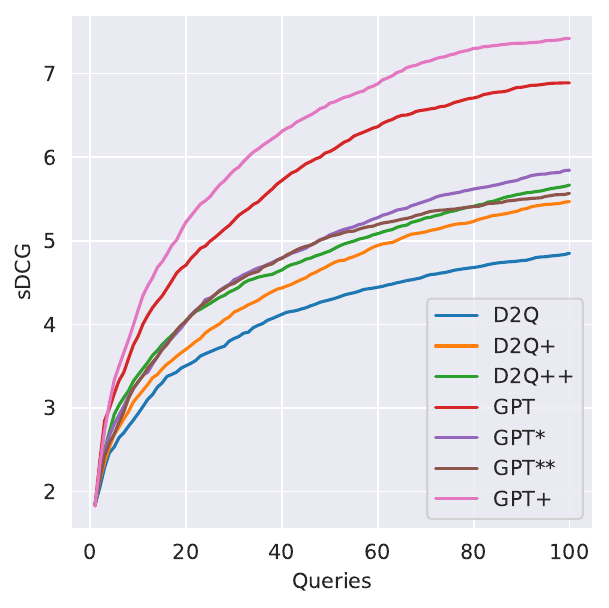}
\includegraphics[width=.325\textwidth]{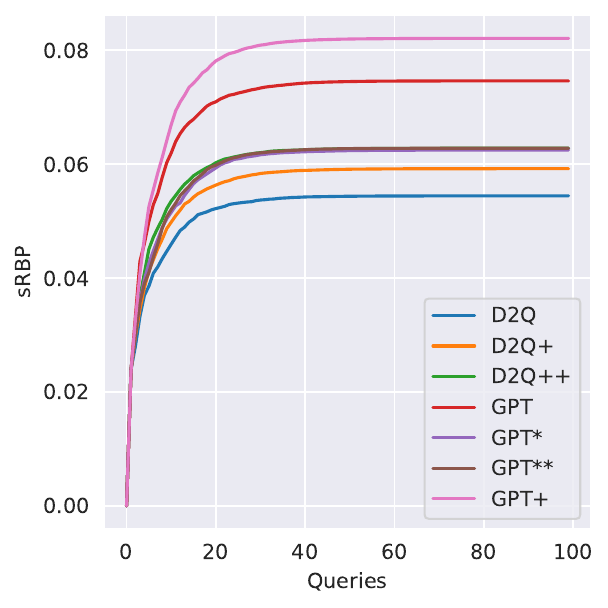}

\caption{Results of the simulated retrieval sessions for different query generation methods. Core18 (top) and Core17 (bottom).}
\label{fig:query_strats}
\end{figure*}

Since it has been shown in Section~\ref{subsec:dense_sparse_eval} that MonoT5 does not cause any substantial differences in the trends of the curves, we continue with only BM25 due to the enormous computational effort.
We examine the following aspects: query generation type, topic background, and feedback (cf. Table~\ref{tab:query_methods}). The plots of the seven different variants are shown in Figure~\ref{fig:query_strats}.

\textbf{Query generation type}
To determine what effect the query variation procedure has, we distinguish between probabilistic and rule-based query variations as described in Section~\ref{sec:query_generation}.
In our case, \textit{GPT} and \textit{GPT+} stand for probabilistic variation, while the rest works ruled-based.
In Figure~\ref{fig:query_strats}, it can be clearly seen that the probabilistic variations have a strong positive effect on the information gain. 
All measures indicate a clear dominance of \textit{GPT} and \textit{GPT+}.

\textbf{Topic background}
To examine the effect of using topic background, we compare the following strategies in pairs.
\textit{GPT} vs. \textit{GPT+}, \textit{GPT*} vs. \textit{GPT**} and \textit{D2Q+} vs. \textit{D2Q++}.
Across all measures, the topic background has a strong positive effect on the probabilistic methods.
Considering the rule-based methods (\textit{GPT*} vs. \textit{GPT**} and \textit{D2Q++} vs. \textit{D2Q+}), the topic background does not substantially affect the information gain.

\textbf{Feedback}
The effect of feedback on information gain is evaluated in our setting by comparing the mechanisms \textit{D2Q+} and \textit{D2Q}.
As described in Section~\ref{sec:query_generation}, \textit{D2Q+} integrates the relevance feedback of the simulated user and accordingly favors terms that come from relevant documents. Comparing the plots displays that feedback leads to a clear increase in information gain. 

\section{Discussion}
The results in Section~\ref{subsec:dense_sparse_eval} show that although the dense retrieval model gives slightly better results, these differences stay the same throughout the simulation.
Minor differences in trends align with other studies~\cite{DBLP:conf/sigir/HershTPCKSO00,DBLP:conf/sigir/TurpinH01}, and in our perspective, user behavior leads to a much greater variance in effectiveness. In particular, taking into account the context within the search session --- modeled by dynamic click behavior, background knowledge, and the integration of feedback --- substantially impacts information gain.

From our point of view, it is crucial for future simulation experiments to integrate the user context instead of only investigating cost-intensive retrieval models. For both test collections, the trends were similar in all evaluations, even though the absolute information gain differed. We suspect this is related to the substantial difference in the number of documents rated as relevant (cf.~\ref{sec:implementation}). 

In future evaluations, the costs for specific user actions used for the effort-based measure should be specified and justified.
Still, the order and trends of the effort-based evaluations resemble the \ac{sDCG} evaluations closely. Furthermore, the effort-based measure can reveal effects that would remain hidden if only the number of queries were considered. 
For example, Figure~\ref{fig:experimental_results_stop_decs} shows that the difference between 50tnr and 20rpp is more pronounced for the effort-based measure since the larger number of examined documents comes with high costs for the user that sDCG does not consider.  Regarding sRBP, sDCG also mainly supports the produced orders of the different configurations. However, the information gain stagnates more quickly, and a difference in the orders can be seen in Figure~\ref{fig:experimental_results_stop_decs}, so dynamic click behavior is penalized by sRBP. Both findings can be explained by the substantially stronger discounting of late results by sRBP. Interestingly, dynamic stop behavior is notably rewarded by the other measures. As Figure~\ref{fig:snippet_dist} indicates, the larger number of examined documents can likely further be attributed to this opposing evaluation behavior. Determining not only realistic durations for user actions but also estimating suitable parameters for sRBP is vital for future work to enable more accurate evaluations and could be done by analyzing real user behavior. 

The use of probabilistic query generation methods shows promising results. Not only do they dominate rule-based approaches, but they are also more capable of integrating background knowledge of the topic.
Integrating feedback into probabilistic approaches would be an exciting direction for future work, as our rule-based approach has clearly shown that feedback is beneficial. Limitations of this work are, on the one hand, a need for estimates of how similar the simulated behavior (e.g., query generation) corresponds to real users. On the other hand, the results obtained cannot be generalized arbitrarily since we only examined a limited variation of simulated users for a specific search task, and both data sets correspond to the same domain.

\section{Conclusion}

In this work, we propose novel ways to simulate context-dependent query formulations that are evaluated with a state-of-the-art experimental setup, including LLMs and dense retrieval methods. To better understand the effects of context information, we compare context-sensitive query simulation methods against context-free variants. Our experimental results suggest that both the inclusion of comprehensive descriptions of the information need as well as the feedback inferred from earlier seen search results impact the progress of a simulation session, leading to better search effectiveness. In this regard, we envision experiments with user simulations of higher fidelity by considering contextual information. \\ \\
\noindent \textbf{Acknowledgements} This work was supported by Klaus Tschira Stiftung (JoIE - 00.003.2020) and Deutsche Forschungsgemeinschaft (RESIRE - 509543643).

\bibliographystyle{splncs04}
\bibliography{bibliography}

\end{document}